\documentclass[conference]{IEEEtran}

\IEEEoverridecommandlockouts
\usepackage{cite}
\usepackage{amsmath,amssymb,amsfonts}
\usepackage{algorithmic}
\usepackage{tikz}
\usetikzlibrary{arrows.meta, positioning}
\usepackage{algorithm}
\usepackage{array}
\usepackage{graphicx}
\usepackage{textcomp}
\usepackage{xcolor}
\usepackage{booktabs}
\usepackage{tabularx}
\usepackage{pifont}
\usepackage{float}      
\usepackage{placeins} 
\usepackage[caption=false,font=footnotesize]{subfig}
\usepackage{fancyhdr}
\fancypagestyle{acceptednotice}{
    \fancyhf{}
    
    \fancyhead[C]{\textcolor{red}{\textbf{Accepted for presentation at the IEEE INFOCOM 2026 Workshop (ISAC-FutureG), Tokyo, Japan}}}
}

\newcommand{\cmark}{\ding{51}}
\newcommand{\xmark}{\ding{55}}
\def\BibTeX{{\rm B\kern-.05em{\sc i\kern-.025em b}\kern-.08em
    T\kern-.1667em\lower.7ex\hbox{E}\kern-.125emX}}
\begin{document}

\title{Digital Twin-Assisted Belief-State Reinforcement Learning for Latency-Robust ISAC in 6G Networks}

\author{
\IEEEauthorblockN{Himanshu Tiwari\IEEEauthorrefmark{1}\IEEEauthorrefmark{2}, Binayak Kar\IEEEauthorrefmark{1}\IEEEauthorrefmark{2}, and Priyanshu Tiwari\IEEEauthorrefmark{3}}
\thanks{This work was supported by the National Science and Technology Council, Taiwan, under Grant 114-2221-E-011-137-}
\\ 
\IEEEauthorblockA{\small \IEEEauthorrefmark{1}Department of Computer Science and Information Engineering, National Taiwan University of Science and Technology, Taipei, Taiwan}
\IEEEauthorblockA{\small \IEEEauthorrefmark{2}Quantum Research Lab, National Taiwan University of Science and Technology, Taipei, Taiwan}
\IEEEauthorblockA{\small \IEEEauthorrefmark{3}Department of Artificial Intelligence and Machine Learning, Sir M. Visvesvaraya Institute of Technology, Bengaluru, India}
Email: nomails1337@gmail.com, bkar@mail.ntust.edu.tw, techarena955@gmail.com
}

\maketitle
\thispagestyle{acceptednotice}

\begin{abstract}
Integrated Sensing and Communication (ISAC) enables joint data transmission and environmental perception for sixth-generation (6G) networks, but centralized and virtualized RAN control loops introduce telemetry latency that yields stale observations and unstable control. This paper proposes a Digital Twin-assisted belief-state reinforcement learning framework for latency-robust ISAC. A Digital Twin (DT) reconstructs a synchronized belief state from delayed telemetry using an Extended Kalman Filter, and a Proximal Policy Optimization agent performs joint beamforming and power allocation for communication and sensing. Closed-loop simulations with telemetry delays up to 100~ms demonstrate consistent performance gains over latency-unaware deep reinforcement learning (DRL) and heuristic baselines. At 50~ms latency, the proposed method improves median throughput by 12\% and reduces sensing error by 7\% relative to a DT-only controller, while achieving an order-of-magnitude reduction in reliability violations. Even at 100~ms latency, the proposed approach retains approximately 88\% of its zero-latency throughput. These results show that Digital Twin-assisted belief-state control enables stable and efficient ISAC operation under realistic telemetry delays in 6G networks.
\end{abstract}

\begin{IEEEkeywords}
Integrated sensing and communication, digital twin, reinforcement learning, telemetry latency, 6G networks.
\end{IEEEkeywords}

\section{Introduction}
\label{sec:intro}

Sixth-generation (6G) wireless networks are envisioned to natively support cyber-physical applications that demand tight coupling between wireless connectivity, environmental awareness, and real-time control~\cite{giordani2020towards6g}. Integrated Sensing and Communication (ISAC) has emerged as a key enabling paradigm to meet these requirements by jointly supporting data transmission and radar-like sensing using shared spectrum and radio-frequency hardware. By unifying communication and sensing, ISAC enables advanced functionalities such as high-precision localization, target tracking, and environment mapping, which are essential for autonomous systems, smart cities, and immersive applications~\cite{liu2022isacjsac,kaushik2024isac6g}.

The joint optimization of communication and sensing, however, is inherently challenging. Communication objectives favor directional beamforming and power concentration toward user equipment (UEs) to maximize spectral efficiency, whereas sensing performance benefits from spatial beam diversity and wide-area illumination to reduce estimation error, especially in mobile and dynamic environments~\cite{wu2024isac_interference}. This fundamental trade-off has motivated the adoption of learning-based control, particularly deep reinforcement learning (DRL), as a promising alternative to static or model-driven optimization. Among DRL methods, Proximal Policy Optimization (PPO) is especially attractive for ISAC due to its stability, sample efficiency, and ability to handle continuous action spaces, such as joint beam steering and power allocation.

Despite these advantages, applying PPO directly to practical ISAC control faces a critical limitation in centralized and virtualized radio access networks (RANs), including Open RAN (O-RAN) architectures. In such systems, near-real-time controllers rely on telemetry that is subject to transport, processing, and scheduling delays~\cite{polese2024openran}. Practical telemetry latency, often on the order of tens of milliseconds, results in stale observations and partial observability. When a PPO agent maps delayed telemetry directly to control actions, the selected beams and power levels are applied to an already-evolved physical state, leading to beam misalignment, degraded sensing accuracy, and reduced reliability as latency increases~\cite{park2021latency_ran,meng2020pomdp_wireless}. Consequently, latency-unaware DRL policies can become unstable and lose their performance benefits under realistic delay conditions.

To address this challenge, we propose a Digital Twin (DT)-assisted belief-state reinforcement learning framework for latency-robust ISAC control. Rather than operating PPO directly on delayed observations, a Digital Twin maintains a synchronized virtual replica of the physical ISAC environment and reconstructs a current-time belief state from delayed telemetry using an Extended Kalman Filter (EKF)~\cite{yang2023dt_control}. This belief state compensates for telemetry latency by predicting the evolution of channel and geometry states, thereby restoring effective observability at the controller. A PPO agent then operates on this synchronized belief state to generate continuous joint actions for beam steering and communication and sensing power allocation under feasibility constraints. Closed-loop simulations with telemetry delays up to 100~ms demonstrate that belief-state alignment enables stable learning and preserves the ISAC trade-off under realistic delay regimes.

The main contributions of this paper are summarized as follows:
\begin{itemize}
    \item We formulate a latency-aware ISAC control problem for centralized and virtualized 6G RANs, explicitly capturing partial observability induced by telemetry delay.
    \item We integrate a Proximal Policy Optimization (PPO) agent for continuous joint beamforming and power allocation, and identify its limitations under delayed observations.
    \item We design a Digital Twin-assisted belief-state synchronization mechanism using EKF-based prediction and update to reconstruct a current-time state from delayed telemetry.
    \item We demonstrate through extensive simulations that belief-state PPO significantly improves throughput retention, sensing accuracy, power efficiency, and reliability under telemetry delays ranging from $0$ to $100$~ms.
\end{itemize}

The remainder of this paper is organized as follows. Section~\ref{sec:related_work} reviews related work, Section~\ref{sec:methodology} presents the proposed framework, Section~\ref{sec:results} discusses performance evaluation, and Section~\ref{sec:conclusion} concludes the paper.

\section{Related Work}
\label{sec:related_work}

Integrated Sensing and Communication (ISAC) is a core capability envisioned for 6G networks, motivating extensive research on joint communication and sensing resource optimization. Early optimization-driven works primarily focus on joint beamforming and power allocation to balance spectral efficiency and sensing accuracy under static or slowly varying conditions. Xie \textit{et al.}~\cite{xie2025noma_isac} proposed a NOMA-empowered ISAC framework using constrained nonconvex optimization, assuming accurate and timely channel and geometry information at the controller. While effective under ideal feedback, such methods are sensitive to delayed or stale state information.

Mobility-aware ISAC designs further expose the limitations of static optimization. Lyu \textit{et al.}~\cite{lyu2023uav_isac} studied joint maneuver and beamforming for UAV-enabled ISAC systems, demonstrating that communication and sensing objectives can rapidly drift under motion. However, the framework assumes prompt state acquisition and does not explicitly address telemetry delay or partial observability in centralized control loops.

To cope with nonconvexity and fast dynamics, deep reinforcement learning (DRL) has been applied to ISAC resource management. Long \textit{et al.}~\cite{long2024ris_isac_drl} employed DRL for RIS-assisted ISAC in vehicular networks, while Zhu \textit{et al.}~\cite{zhu2023star_ris_isac} extended DRL-based control to STAR-RIS-assisted secure ISAC systems. Although these approaches enable adaptive control without hand-crafted heuristics, they generally assume timely observations and do not reconstruct the current physical state when telemetry is delayed. As a result, their performance degrades in centralized architectures where delayed feedback induces partial observability.

In parallel, Digital Twin (DT) research for 6G and O-RAN has focused on network mirroring, orchestration, and data-driven optimization support. Tao \textit{et al.}~\cite{tao2024dt_generative_ai} explored DT-enabled wireless networks augmented with generative AI for predictive analysis, without instantiating a closed-loop ISAC controller. Sun and To~\cite{sun2025dt_oran} investigated DT-based O-RAN architectures emphasizing synchronization under latency, but without addressing joint communication and sensing optimization.

Delay-aware orchestration mechanisms have also been proposed for O-RAN. ORANUS~\cite{adamuz2024oranus} and MAREA~\cite{adamuz2025marea} introduced latency-aware and multi-timescale radio resource orchestration frameworks that explicitly account for control-plane delays. However, these works neither incorporate an ISAC objective nor employ learning-based policies operating on synchronized belief states.

Overall, existing literature remains fragmented across three directions: ISAC optimization that neglects telemetry latency, DRL-based ISAC methods that do not recover the current state under delay, and DT or O-RAN studies that address latency without jointly optimizing communication and sensing. In contrast, this paper integrates Digital Twin-assisted belief-state synchronization with PPO-based continuous control for joint beamforming and power allocation in delayed closed-loop ISAC, explicitly targeting latency-prone centralized 6G architectures.
Table~\ref{tab:rw_comparison} summarizes representative works, where \textbf{BF} denotes beamforming, \textbf{Pwr} denotes power allocation, and \textbf{Belief} indicates belief-state synchronization.

\begin{table}[!t]
\centering
\caption{Comparison with related works}
\label{tab:rw_comparison}
\scriptsize
\setlength{\tabcolsep}{3pt}
\renewcommand{\arraystretch}{1.15}
\begin{tabular}{|p{2.15cm}|c|c|c|c|c|c|c|c|c|}
\hline
\textbf{Work} &
\textbf{ISAC} &
\textbf{6G} &
\textbf{DT} &
\textbf{DRL} &
\textbf{BF} &
\textbf{Pwr} &
\textbf{Latency} &
\textbf{PPO} &
\textbf{Belief} \\
\hline
Xie \textit{et al.}~\cite{xie2025noma_isac} &
\cmark & \cmark & \xmark & \xmark & \cmark & \cmark & \xmark & \xmark & \xmark \\
\hline
Lyu \textit{et al.}~\cite{lyu2023uav_isac} &
\cmark & \xmark & \xmark & \xmark & \cmark & \xmark & \xmark & \xmark & \xmark \\
\hline
Long \textit{et al.}~\cite{long2024ris_isac_drl} &
\cmark & \cmark & \xmark & \cmark & \xmark & \xmark & \xmark & \xmark & \xmark \\
\hline
Zhu \textit{et al.}~\cite{zhu2023star_ris_isac} &
\cmark & \cmark & \xmark & \cmark & \cmark & \xmark & \xmark & \xmark & \xmark \\
\hline
Tao \textit{et al.}~\cite{tao2024dt_generative_ai} &
\xmark & \cmark & \cmark & \xmark & \xmark & \xmark & \xmark & \xmark & \xmark \\
\hline
Sun and To~\cite{sun2025dt_oran} &
\xmark & \cmark & \cmark & \xmark & \xmark & \xmark & \cmark & \xmark & \xmark \\
\hline
ORANUS~\cite{adamuz2024oranus} &
\xmark & \cmark & \xmark & \xmark & \xmark & \cmark & \cmark & \xmark & \xmark \\
\hline
MAREA~\cite{adamuz2025marea} &
\xmark & \cmark & \xmark & \xmark & \xmark & \cmark & \cmark & \xmark & \xmark \\
\hline
\textbf{Proposed} &
\cmark & \cmark & \cmark & \cmark & \cmark & \cmark & \cmark & \cmark & \cmark \\
\hline
\end{tabular}
\end{table}

\section{Proposed Framework}
\label{sec:methodology}

\begin{figure}[!t]
\centering
\includegraphics[width=\linewidth]{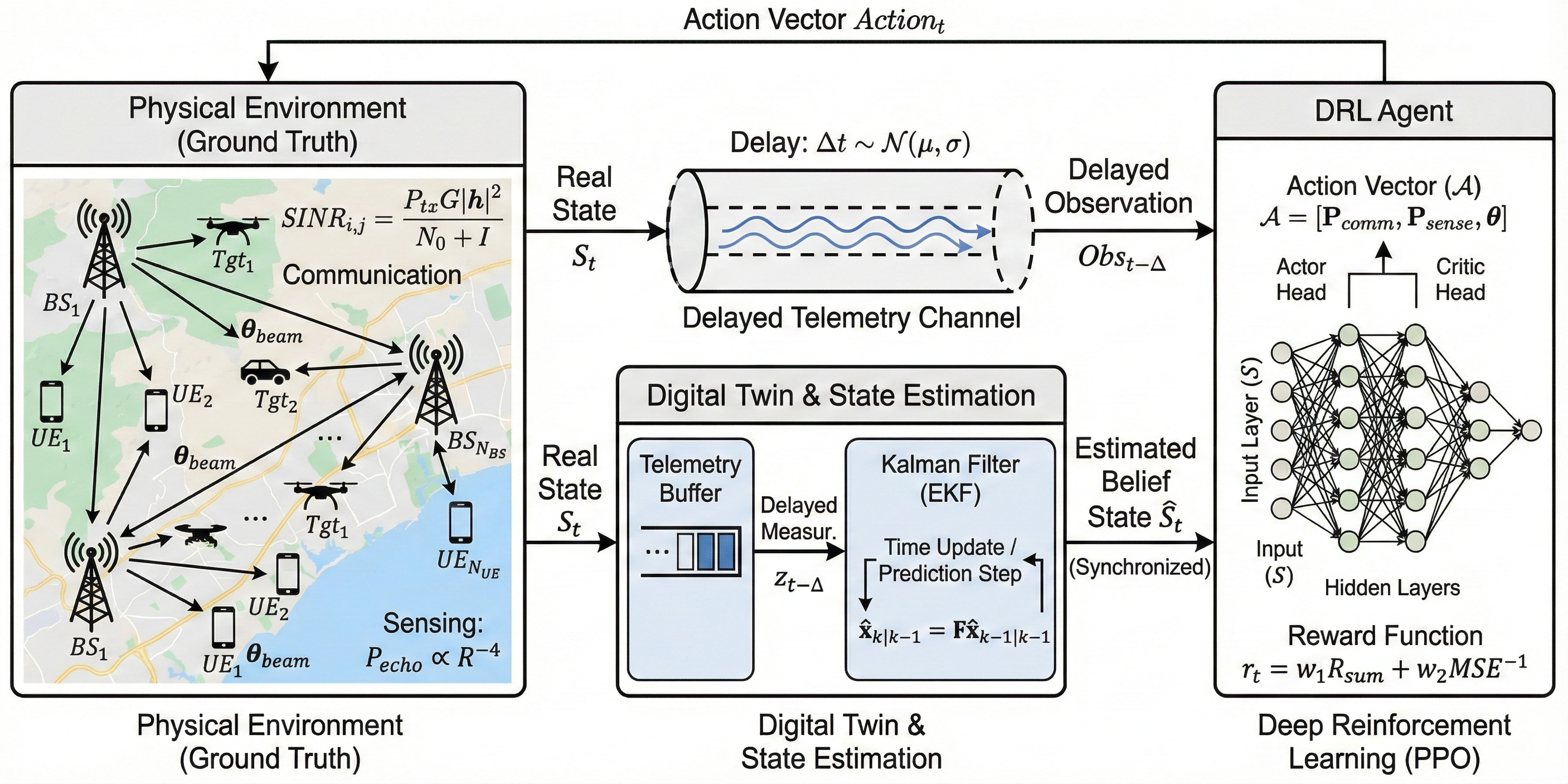}
\caption{Proposed Digital Twin-assisted ISAC framework with delayed telemetry and belief-state control.}
\label{fig:system_arch}
\end{figure}

This section details the proposed Digital Twin-assisted ISAC control loop under delayed telemetry. The framework couples a high-fidelity Digital Twin with belief-state synchronization and a PPO-based actor–critic agent for joint beamforming and power allocation. The physical network evolves according to the ground-truth state $S_t$, while the controller receives delayed measurements $Obs_{t-\Delta}$. The Digital Twin buffers these measurements and reconstructs a synchronized belief state $\hat{S}_t$ using an Extended Kalman Filter (EKF). The PPO policy maps $\hat{S}_t$ to a continuous action vector $\mathcal{A}_t$ that is applied to the physical ISAC environment. The end-to-end architecture is shown in Fig.~\ref{fig:system_arch}.

\subsection{Physical ISAC Environment}

We consider a downlink multi-cell ISAC network with $N_{BS}$ base stations, $N_{UE}$ user equipments, and $N_{Tgt}$ sensing targets. Each base station uses a shared RF front-end for communication and radar sensing, forming beams that simultaneously serve users and probe targets. The ground-truth environment state at time $t$ is denoted by $S_t$, which includes UE and target kinematics, channel and geometry information, and any additional variables needed to evaluate communication and sensing utilities.

\subsubsection{Communication Model}

For UE $j$ served by BS $i$, the received power is
{\small
\begin{equation}
P_{rx}(i,j)=P_{tx}\,G_t(\boldsymbol{\theta}_i,\phi_j)\,G_r\,L(d_{i,j})\,|h_{i,j}|^2
\end{equation}
}

where $P_{tx}$ is the transmit power, $G_t(\cdot)$ is the transmit beamforming gain parameterized by steering angles $\boldsymbol{\theta}_i$, $G_r$ is the receive gain, $L(d_{i,j})=\beta d_{i,j}^{-\alpha}$ is the path-loss model, and $h_{i,j}\sim\mathcal{CN}(0,1)$ represents Rayleigh fading. The achievable rate for UE $j$ is

{\small
\begin{equation}
R_j = B\log_2\!\left(1+\frac{\sum_{i\in\mathcal{C}}P_{rx}(i,j)}{\sigma^2+I_{\text{inter}}}\right),
\end{equation}
}

where $B$ is bandwidth, $\sigma^2$ is noise power, and $I_{\text{inter}}$ denotes inter-cell interference. The communication utility is the sum rate

{\small
\begin{equation}
R_{\text{sum}}=\sum_{j=1}^{N_{UE}}R_j.
\end{equation}
}
\subsubsection{Sensing Model}

For sensing, base stations transmit probing waveforms toward targets. The received echo power from target $k$ at range $R_k$ satisfies the radar range equation

{\small
\begin{equation}
P_{\text{echo},k} \propto \frac{P_{\text{sense}}\,G_{tx}(\theta_k)\,G_{rx}(\theta_k)\,\sigma_{RCS,k}}{R_k^4},
\end{equation}
}
where $P_{\text{sense}}$ is sensing power, $G_{tx}(\cdot)$ and $G_{rx}(\cdot)$ are gains in the target direction, and $\sigma_{RCS,k}$ is the radar cross section. Sensing performance is quantified by the target position estimation mean squared error

{\small
\begin{equation}
\text{MSE} = \frac{1}{N_{Tgt}}\sum_{k=1}^{N_{Tgt}}\|\hat{\mathbf{p}}_{k,t}-\mathbf{p}_{k,t}\|_2^2,
\end{equation}
}
where $\mathbf{p}_{k,t}$ is the ground-truth target position and $\hat{\mathbf{p}}_{k,t}$ is the corresponding estimate.

\subsection{Delayed Telemetry Channel}

In centralized control, the controller does not observe $S_t$ directly. Instead, it receives a delayed observation $Obs_{t-\Delta}$ through a telemetry channel with random delay. We model latency as
{\small
\begin{equation}
\Delta t \sim \mathcal{N}(\mu,\sigma^2),
\end{equation}
}
so that telemetry arriving at decision time $t$ corresponds to an earlier physical state:
{\small
\begin{equation}
Obs_{t-\Delta} = \mathcal{H}(S_{t-\Delta}) + \boldsymbol{\nu}_{t-\Delta},
\end{equation}
}
where $\mathcal{H}(\cdot)$ is the measurement function and $\boldsymbol{\nu}_{t-\Delta}$ is measurement noise. This delay creates partial observability and can destabilize beamforming and power control when system dynamics evolve faster than the feedback loop.
Since physical latency cannot be negative, delays are truncated to enforce $\Delta t \ge 0$. In implementation, we apply $\Delta t \leftarrow \max(\Delta t,0)$, and we quantize $\Delta t$ to the simulation step to ensure consistent alignment between measurement timestamps and control instants. This models realistic controller-side buffering and scheduling effects in centralized RAN deployments.

\subsection{Digital Twin and Belief-State Synchronization}

The Digital Twin compensates telemetry delay by buffering time-stamped measurements and reconstructing a synchronized belief state aligned to the current decision instant. Let $\mathcal{B}_t$ denote a finite telemetry buffer:
{\small
\begin{equation}
\mathcal{B}_t=\{(z_{\tau},\tau)\mid \tau \le t\},
\end{equation}
}
where $z_{\tau}$ is a measurement generated at time $\tau$. At each control step, the Digital Twin selects the newest available delayed measurement that has arrived by time $t$, denoted $z_{t-\Delta}$, and performs EKF-based prediction and update to synchronize the state.

\subsubsection{EKF State and Dynamics}

For each tracked entity (UE or target), the EKF state vector is
{\small
\begin{equation}
\mathbf{x}_t=[p_x,\,p_y,\,v_x,\,v_y]^T.
\end{equation}
}
Under a constant-velocity model with sampling interval $\Delta t$, state evolution is
{\small
\begin{equation}
\mathbf{x}_{t}=\mathbf{F}\mathbf{x}_{t-1}+\mathbf{w}_{t-1}, \quad \mathbf{w}_{t-1}\sim\mathcal{N}(\mathbf{0},\mathbf{Q}),
\end{equation}
}
where
{\small
\begin{equation}
\mathbf{F}=
\begin{bmatrix}
1&0&\Delta t&0\\
0&1&0&\Delta t\\
0&0&1&0\\
0&0&0&1
\end{bmatrix}.
\end{equation}
}
Measurements satisfy
{\small
\begin{equation}
\mathbf{z}_{t-\Delta} = h(\mathbf{x}_{t-\Delta}) + \mathbf{v}_{t-\Delta}, \quad \mathbf{v}_{t-\Delta}\sim\mathcal{N}(\mathbf{0},\mathbf{R}),
\end{equation}
}
where $h(\cdot)$ maps latent kinematics and geometry to telemetry measurements. The EKF applies standard linearization with Jacobians to propagate covariance and correct the posterior estimate.

\subsubsection{Belief-State Construction}

The synchronized belief state $\hat{S}_t$ aggregates EKF outputs across all UEs and targets, together with derived link-quality indicators, such as predicted CSI or SINR proxies:
{\small
\begin{equation}
\hat{S}_t = \Big[\{\hat{\mathbf{x}}^{UE}_{j,t}\}_{j=1}^{N_{UE}},\{\hat{\mathbf{x}}^{Tgt}_{k,t}\}_{k=1}^{N_{Tgt}}, \widehat{\text{CSI}}_t \Big].
\end{equation}
}
This belief-state synchronization reduces the effective observation delay seen by the controller, enabling stable policy execution under telemetry latency.

\subsection{PPO Agent for Joint ISAC Control}

The PPO agent operates on $\hat{S}_t$ and outputs a continuous action vector
{\small
\begin{equation}
\mathcal{A}_t = [P_{\text{comm},t},\,P_{\text{sense},t},\,\boldsymbol{\theta}_t],
\end{equation}
}
where $P_{\text{comm},t}$ and $P_{\text{sense},t}$ allocate power to communication and sensing, and $\boldsymbol{\theta}_t$ denotes beam steering parameters. Feasibility is enforced through
{\small
\begin{equation}
P_{\text{comm},t}\ge 0,\quad P_{\text{sense},t}\ge 0,\quad P_{\text{comm},t}+P_{\text{sense},t}\le P_{\max}.
\end{equation}
}
The reward balances throughput, sensing accuracy, and power discipline:
{\small
\begin{equation}
r_t = w_1 R_{\text{sum},t} + w_2 \left(\text{MSE}_t\right)^{-1} - w_3\frac{P_{\text{comm},t}+P_{\text{sense},t}}{P_{\max}},
\end{equation}
}
where $w_1,w_2,w_3$ control the operating point on the ISAC trade-off surface. The PPO agent uses an actor–critic architecture, where the actor parameterizes $\pi_{\theta}(\mathcal{A}_t\mid\hat{S}_t)$ and the critic estimates $V_{\psi}(\hat{S}_t)$, enabling stable updates via the PPO clipped objective.

\section{Algorithm and Experimental Results}
\label{sec:results}

This section evaluates the proposed Digital Twin-assisted belief-state control with PPO (DT + EKF + PPO) under telemetry latency using the closed-loop architecture in Fig.~\ref{fig:system_arch}. Performance is assessed via communication throughput, sensing accuracy, power efficiency, reliability (constraint violations), and the overall ISAC objective. All results are obtained from closed-loop simulations with delayed telemetry, and medians are reported unless stated otherwise.

\subsection{Training, Deployment, and Simulation Setup}
\label{subsec:train_deploy_sim}

As illustrated in Fig.~\ref{fig:system_arch}, the controller does not act directly on delayed telemetry. Instead, the Digital Twin buffers time-stamped observations and reconstructs a synchronized belief state $\hat{S}_t$ via EKF prediction and update. Algorithm~\ref{alg:dt_ppo} formalizes this process. At each timestep, the PPO actor maps $\hat{S}_t$ to a continuous joint action $\mathcal{A}_t=[P_{\text{comm},t},P_{\text{sense},t},\boldsymbol{\theta}_t]$, while feasibility is enforced through non-negativity and total power constraints.

During training, interactions generated by Algorithm~\ref{alg:dt_ppo} are collected and PPO is updated using the clipped surrogate objective, ensuring stable learning under partial observability. During deployment, the same algorithm runs online, with the Digital Twin continuously converting delayed telemetry into belief states, thereby aligning control actions with the predicted current physical state rather than stale measurements.

\begin{algorithm}[!t]
\caption{Digital Twin-assisted PPO for ISAC with delayed telemetry}
\label{alg:dt_ppo}
{\scriptsize
\begin{algorithmic}[1]
\STATE Initialize actor $\pi_{\theta}$ and critic $V_{\psi}$
\STATE Initialize telemetry buffer $\mathcal{B}$ and EKF parameters $(\mathbf{Q},\mathbf{R})$
\FOR{each episode}
    \STATE Reset physical environment and Digital Twin
    \FOR{each timestep $t$}
        \STATE Receive delayed observation $Obs_{t-\Delta}$ and push into $\mathcal{B}_t$
        \STATE Select newest available delayed measurement $z_{t-\Delta}$ from $\mathcal{B}_t$
        \STATE EKF prediction and update $\Rightarrow$ synchronized belief state $\hat{S}_t$
        \STATE Sample $\mathcal{A}_t \sim \pi_{\theta}(\cdot\mid\hat{S}_t)$ and enforce constraints
        \STATE Apply $\mathcal{A}_t$ to the physical environment
        \STATE Compute $R_{\text{sum},t}$, $\text{MSE}_t$, and reward $r_t$
        \STATE Store $(\hat{S}_t,\mathcal{A}_t,r_t,\hat{S}_{t+1})$
    \ENDFOR
    \STATE Update $\pi_{\theta}$ and $V_{\psi}$ using PPO clipped objective
\ENDFOR
\end{algorithmic}
}
\end{algorithm}

Table~\ref{tab:sim_params} summarizes the simulation parameters used throughout the evaluation. Telemetry latency is swept from $0$ to $100$~ms, consistent with centralized and virtualized RAN control loops, and all baselines share identical channel, mobility, and power settings.

\begin{table}[H]
\centering
\caption{Key simulation parameters.}
\label{tab:sim_params}
{\scriptsize
\setlength{\tabcolsep}{4pt}
\renewcommand{\arraystretch}{1.15}
\begin{tabular}{|l|l|}
\hline
\textbf{Parameter} & \textbf{Value} \\
\hline
Number of BSs $N_{BS}$ & $3$ \\
Number of UEs $N_{UE}$ & $6$ \\
Number of targets $N_{Tgt}$ & $3$ \\
Bandwidth $B$ & $20$~MHz \\
Noise power $\sigma^2$ & $-174$~dBm/Hz \\
Path-loss exponent $\alpha$ & $3.2$ \\
Path-loss constant $\beta$ & $-30$~dB at 1~m \\
Max transmit power $P_{\max}$ & $30$~dBm \\
Mobility model and speed range & Constant-velocity, $0$--$20$~m/s \\
Channel model & Rayleigh fading, $h\sim\mathcal{CN}(0,1)$ \\
Telemetry delay distribution & $\Delta t\sim\mathcal{N}(\mu,\sigma^2)$, truncated to $\Delta t\ge 0$ \\
Mean delay $\mu$ & $50$~ms \\
Delay standard deviation $\sigma$ & $15$~ms \\
Delay sweep range & $0$--$100$~ms \\
\hline
\end{tabular}
}
\end{table}

\subsection{Evaluation Setup and Baselines}
\label{subsec:baselines}

We evaluate a centralized ISAC controller that selects $\mathcal{A}_t$ from delayed telemetry. The proposed DT + EKF + PPO is compared against four baselines: (i) latency-unaware DRL ISAC, (ii) PPO operating directly on delayed observations, (iii) DT-only heuristic control, and (iv) convex or heuristic ISAC without DT or DRL. The key difference is that only the proposed method uses Fig.~\ref{fig:system_arch} to reconstruct a synchronized belief state before taking actions. This removes most of the effective delay seen by the policy, reducing beam misalignment and stabilizing the communication-sensing trade-off as $\Delta$ grows.

\subsection{Pareto Trade-Off at 50~ms Latency}
\label{subsec:pareto}

\begin{figure}[H]
\centering
\includegraphics[width=0.7\linewidth]{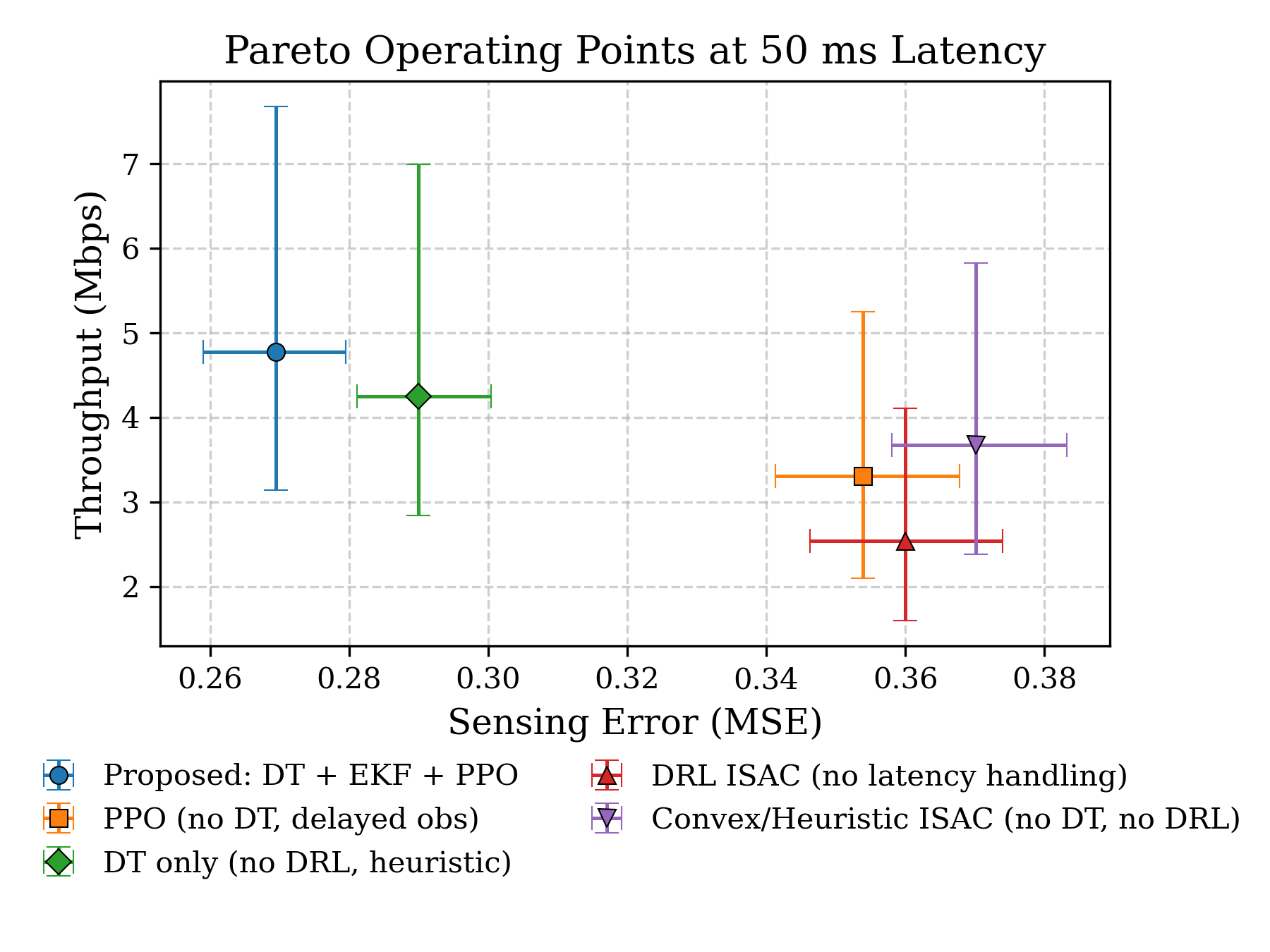}
\caption{Pareto operating points at 50~ms telemetry latency. Higher throughput and lower sensing MSE are desirable.}
\label{fig:pareto_50ms}
\end{figure}

Fig.~\ref{fig:pareto_50ms} shows throughput versus sensing MSE at $50$~ms latency. The proposed method achieves the best operating region by improving both objectives concurrently. This is because EKF-based belief alignment provides a near current-time estimate of channel and geometry, so PPO can steer beams and split power according to the present state, not a stale one. Consequently, the controller avoids the common failure mode of delayed policies, namely allocating power to misaligned beams (hurting throughput) or widening beams excessively to hedge against uncertainty (hurting sensing and power efficiency). At 50~ms, DT + EKF + PPO improves median throughput by about $12\%$ over DT-only and by more than $40\%$ over delayed-observation PPO, while reducing sensing MSE by about $7\%$ and $24\%$, respectively.

\subsection{Throughput Retention Versus Latency}
\label{subsec:throughput}

\begin{figure}[!t]
\centering
\includegraphics[width=0.7\linewidth]{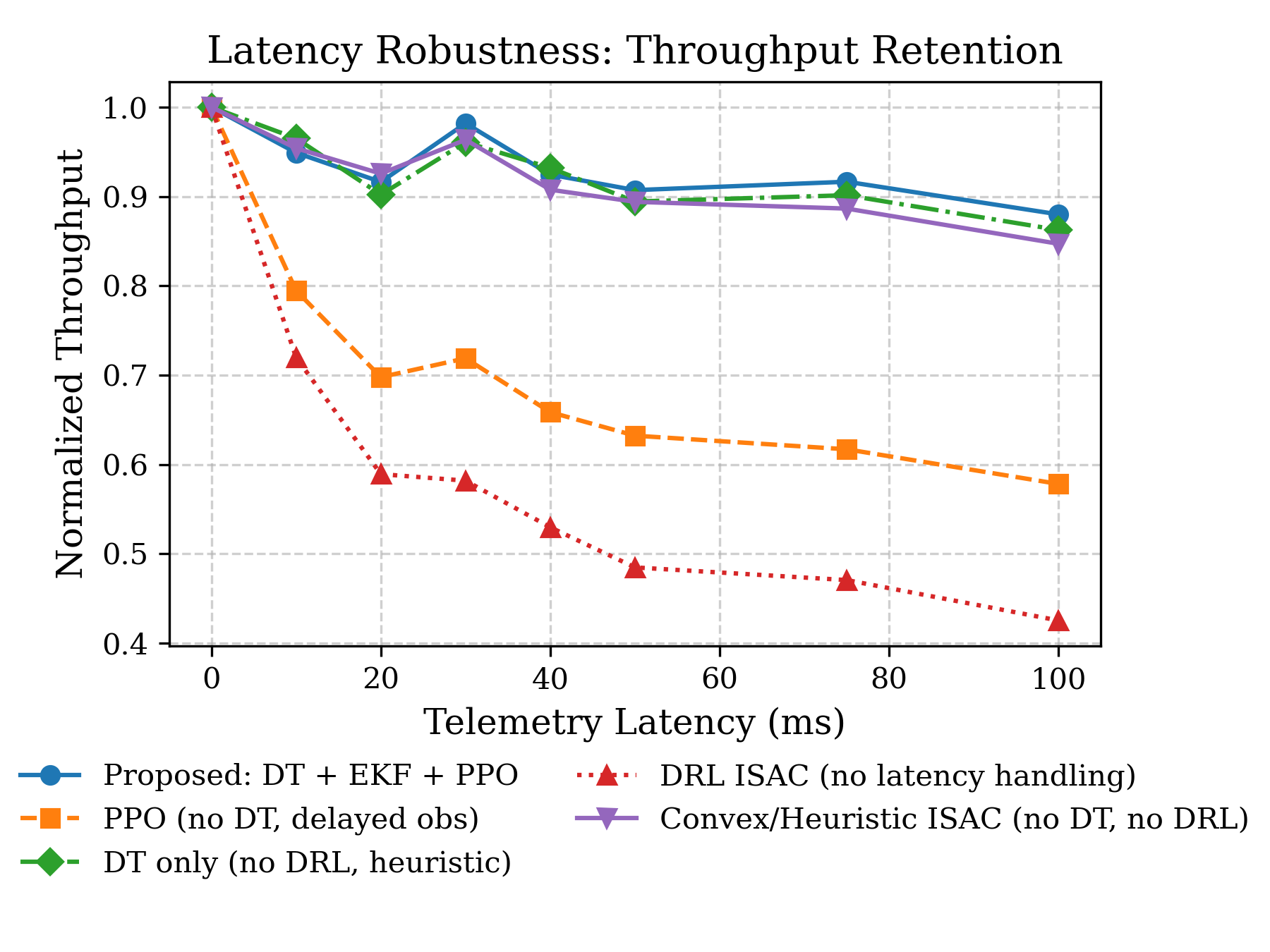}
\caption{Normalized throughput retention versus telemetry latency. Higher values indicate stronger robustness.}
\label{fig:tp_retention}
\end{figure}

Fig.~\ref{fig:tp_retention} reports normalized throughput retention versus telemetry latency. DT + EKF + PPO retains the highest fraction of its zero-latency throughput because the Digital Twin predicts the state forward to the decision time (Fig.~\ref{fig:system_arch}), so beam steering remains aligned with UE motion and channel evolution even when telemetry is delayed. In contrast, delayed-observation PPO and latency-unaware DRL apply actions optimized for past states, which increasingly misalign as latency grows, reducing SINR and sum rate. At 100~ms, the proposed method retains $\approx 88\%$ of its zero-latency throughput, compared to $86\%$ (DT-only), $58\%$ (delayed-observation PPO), and $43\%$ (latency-unaware DRL).

\subsection{Sensing Robustness Versus Latency}
\label{subsec:sensing}

\begin{figure}[H]
\centering
\includegraphics[width=0.7\linewidth]{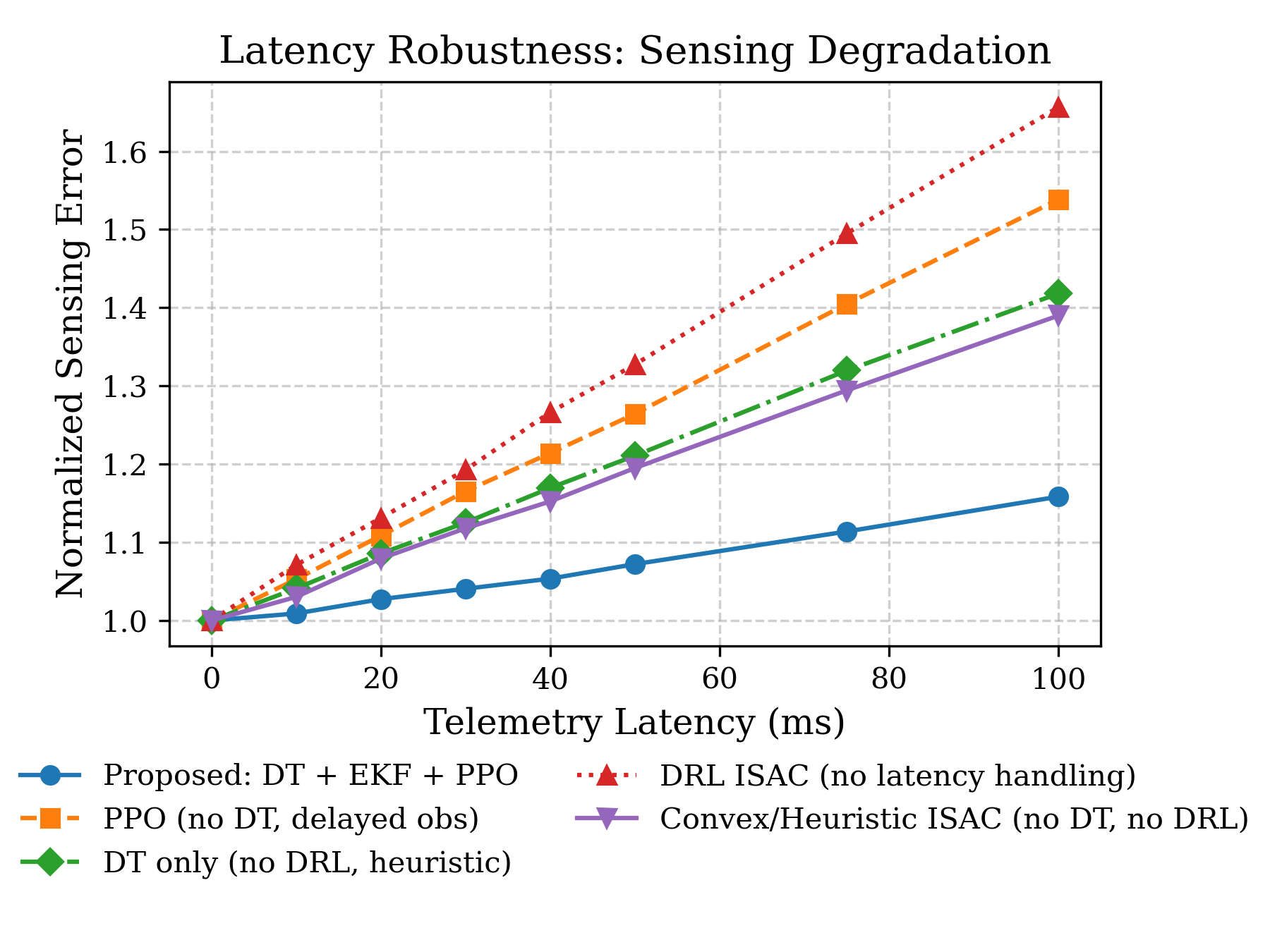}
\caption{Normalized sensing error versus telemetry latency. Lower values indicate better sensing robustness.}
\label{fig:sensing_degradation}
\end{figure}

Fig.~\ref{fig:sensing_degradation} shows normalized sensing error versus telemetry latency. DT + EKF + PPO degrades slowest since EKF prediction preserves target geometry estimates at the current time, enabling beams that illuminate targets with higher effective gain and more consistent tracking quality. Baselines that act on stale measurements suffer geometry drift, which increases localization error and MSE. At 100~ms, the proposed approach reduces median sensing MSE by about $14\%$ relative to DT-only and by more than $35\%$ relative to latency-unaware DRL.

\subsection{Power Efficiency Under Telemetry Delay}
\label{subsec:power}

\begin{figure}[H]
\centering
\includegraphics[width=0.7\linewidth]{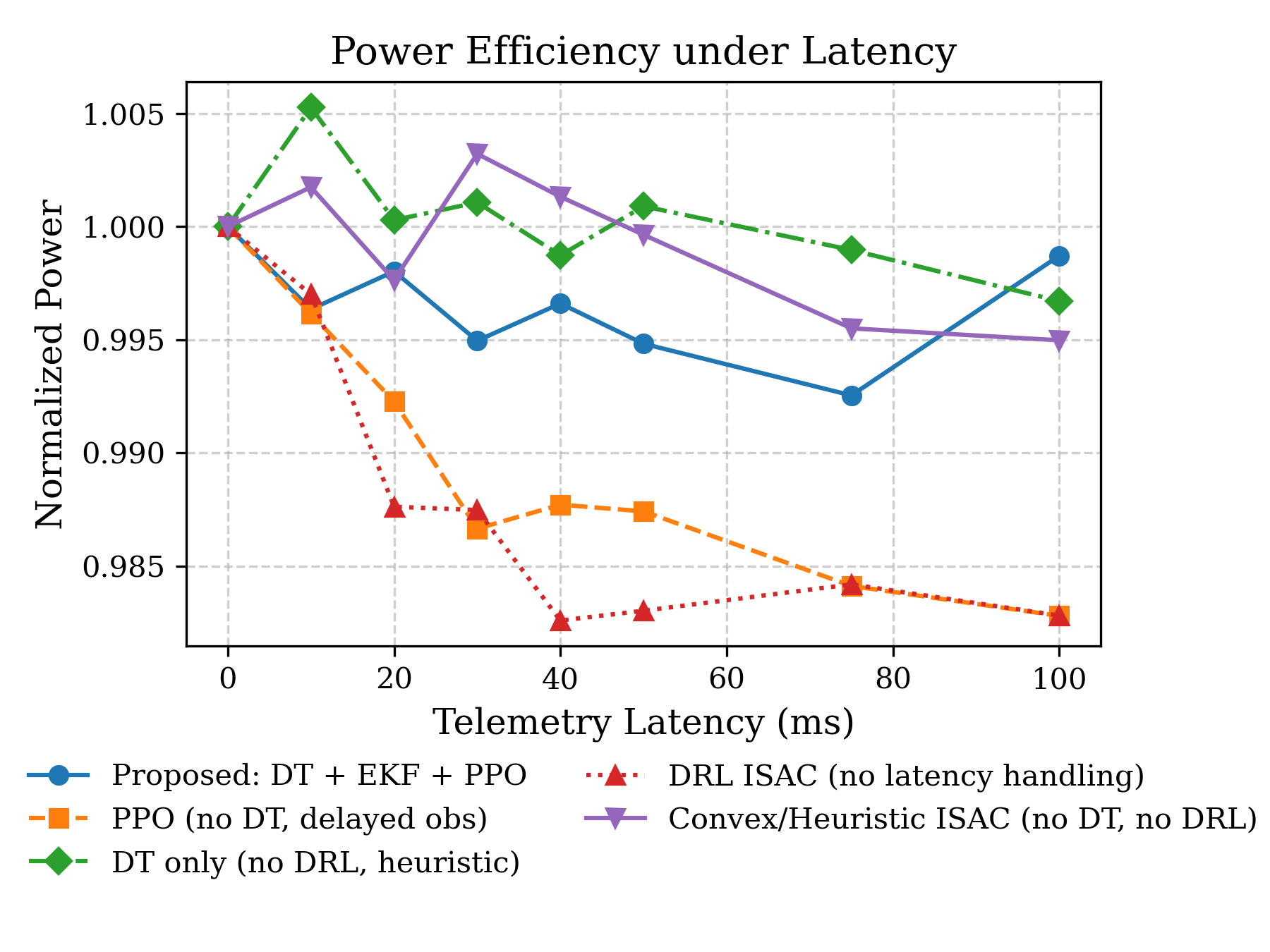}
\caption{Normalized power usage versus telemetry latency. Lower values indicate higher efficiency.}
\label{fig:power_efficiency}
\end{figure}

Fig.~\ref{fig:power_efficiency} reports normalized power usage versus latency. DT + EKF + PPO maintains near-constant power consumption because belief alignment reduces the need for ``power over-compensation'' that occurs when the controller is uncertain due to stale telemetry. With a synchronized $\hat{S}_t$, PPO can allocate the minimum power required to meet communication and sensing goals, instead of increasing power to hedge against beam misalignment. At 50~ms and 100~ms, DT + EKF + PPO reduces median power usage by about $11\%$ compared to DT-only, while delayed-observation policies consume more power with weaker performance gains.

\subsection{Reliability Under Telemetry Latency}
\label{subsec:reliability}

\begin{figure}[H]
\centering
\includegraphics[width=0.7\linewidth]{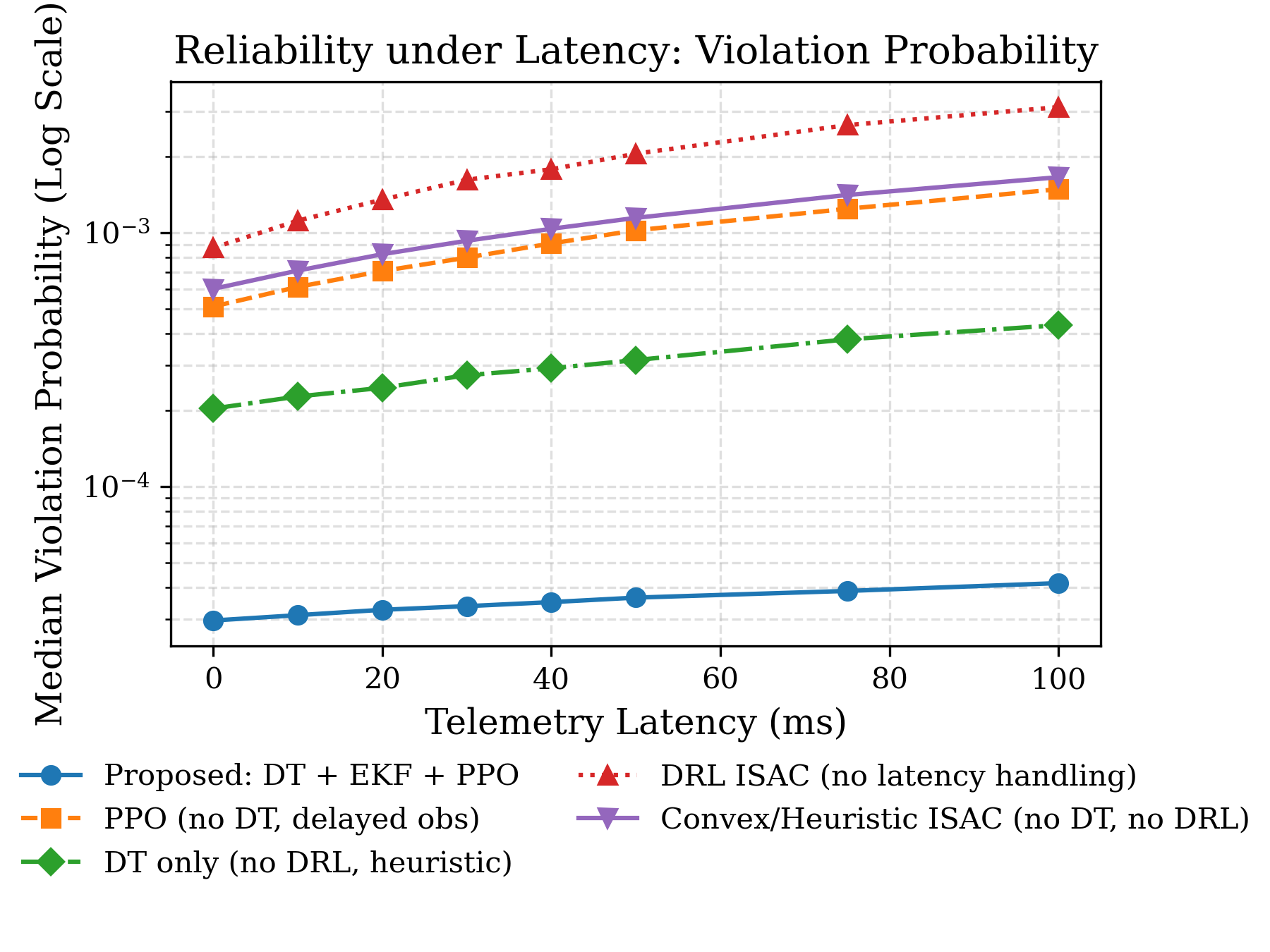}
\caption{Median violation probability versus telemetry latency (log scale). Lower values indicate higher reliability.}
\label{fig:violation_prob}
\end{figure}

Fig.~\ref{fig:violation_prob} shows median violation probability on a logarithmic scale. DT + EKF + PPO achieves the lowest violation probability because synchronized belief states reduce abrupt action swings caused by delayed, inconsistent observations. This stabilizes the closed-loop controller and prevents entering unsafe regimes that trigger reliability violations. DT-only improves over pure delayed DRL by prediction, but it lacks policy learning to optimally manage the ISAC trade-off under uncertainty, yielding higher residual violations. At 50~ms, the proposed method reduces violation probability by nearly one order of magnitude relative to DT-only and by more than two orders of magnitude relative to latency-unaware DRL, with similar trends at 100~ms.

\subsection{Overall ISAC Objective Versus Latency}
\label{subsec:objective}

\begin{figure}[H]
\centering
\includegraphics[width=0.7\linewidth]{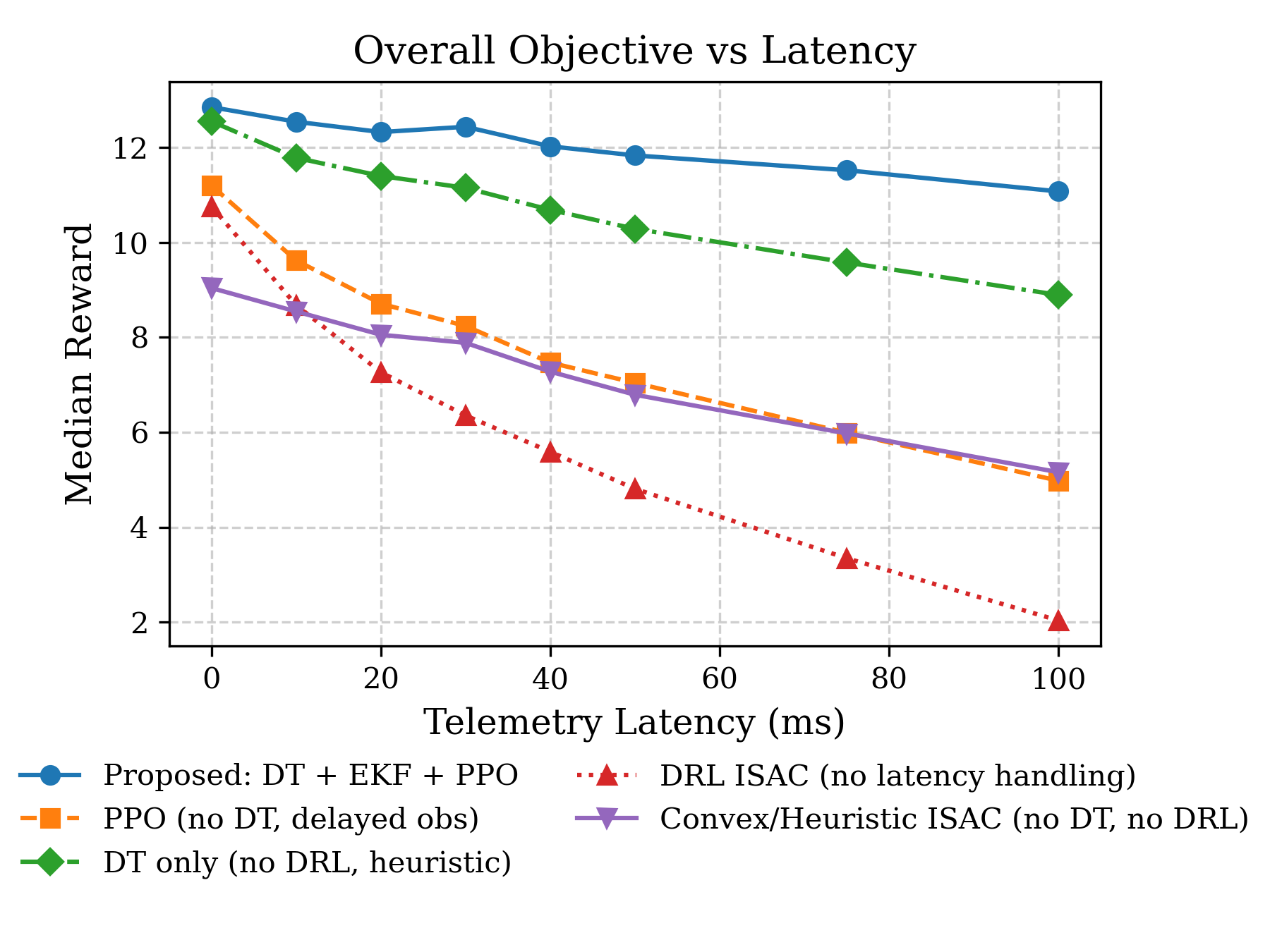}
\caption{Median overall objective versus telemetry latency. Higher values indicate better ISAC trade-off.}
\label{fig:reward_vs_latency}
\end{figure}

Fig.~\ref{fig:reward_vs_latency} reports the median overall ISAC objective. DT + EKF + PPO achieves the highest reward across all latency values because it simultaneously (i) preserves throughput via beam alignment, (ii) limits sensing error via predicted geometry, and (iii) avoids unnecessary power increases and constraint violations. Baselines typically optimize one aspect at the expense of another under delay, for example boosting power to recover throughput while increasing violations, or widening beams to stabilize sensing while losing spectral efficiency. At 50~ms, the proposed method improves median reward by about $15\%$ over DT-only, and at 100~ms the gain increases to more than $24\%$, indicating widening advantages as latency becomes more severe.

\section{Conclusion and Future Work}
\label{sec:conclusion}

This paper addressed latency-robust ISAC control in centralized and virtualized 6G RANs by proposing a Digital Twin-assisted belief-state reinforcement learning framework. By reconstructing synchronized belief states from delayed telemetry using EKF and optimizing joint beamforming and power allocation via PPO, the proposed DT + EKF + PPO consistently outperforms latency-unaware DRL and heuristic baselines, achieving superior throughput-sensing trade-offs, higher power efficiency, and significantly improved reliability under delays up to 100~ms.

Future work will focus on extending the Digital Twin to model non-stationary channels and complex sensing scenarios, and on scaling the framework to larger multi-cell systems using hierarchical or multi-agent RL aligned with O-RAN control timescales.

\bibliographystyle{IEEEtran}
\bibliography{references}

@article{xie2025noma_isac,
  author  = {Xie, Yiran and Yau, David K. Y. and Cheng, Ning and Li, Yu and Aldubaikhy, Khaled},
  title   = {Joint Beamforming and Power Allocation Strategy for {NOMA} Empowered {ISAC} Systems},
  journal = {IEEE Transactions on Vehicular Technology},
  year    = {2025},
  volume  = {74},
  number  = {2},
  pages   = {24205--24219}
}

@article{lyu2023uav_isac,
  author  = {Lyu, Zhonghao and Zhu, Guangxu and Xu, Jie},
  title   = {Joint Maneuver and Beamforming Design for {UAV}-Enabled Integrated Sensing and Communication},
  journal = {IEEE Transactions on Wireless Communications},
  year    = {2023},
  volume  = {22},
  number  = {4},
  pages   = {2424--2440}
}

@article{long2024ris_isac_drl,
  author  = {Long, Hao and Chen, Shaohui and Zeng, Yijian and Xia, Binbin and Nie, Zaiqiang and Xu, Wei and Huang, Yongming},
  title   = {Deep Reinforcement Learning for Integrated Sensing and Communication in {RIS}-Assisted {6G} {V2X} System},
  journal = {IEEE Internet of Things Journal},
  year    = {2024},
  volume  = {11},
  number  = {24},
  pages   = {40691--40703}
}

@inproceedings{zhu2023star_ris_isac,
  author    = {Zhu, Zhengyu and Gong, Mengfei and Chu, Zheng and Xiao, Pei and Sun, Gangcan and Mi, De and He, Ziming and Tong, Fei},
  title     = {{DRL}-based {STAR-RIS}-Assisted {ISAC} Secure Communications},
  booktitle = {Proc. International Conference on Ubiquitous Communication (UCom)},
  year      = {2023},
  doi       = {10.1109/Ucom59132.2023.10257639},
  url       = {https://ieeexplore.ieee.org/document/10257639}
}

@article{tao2024dt_generative_ai,
  author  = {Tao, Zhenyu and Xu, Wei and Huang, Yongming and Wang, Xiaoyun and You, Xiaohu},
  title   = {Wireless Network Digital Twin for {6G}: Generative {AI} as a Key Enabler},
  journal = {IEEE Wireless Communications},
  year    = {2024},
  volume  = {31},
  number  = {4},
  pages   = {24--31}
}

@article{sun2025dt_oran,
  author  = {Sun, Kexuan and To, Duc},
  title   = {Digital Twin for {O-RAN} Towards {6G}},
  journal = {IEEE Communications Magazine},
  year    = {2025},
  volume  = {63},
  number  = {3},
  pages   = {174--181}
}

@inproceedings{adamuz2024oranus,
  author    = {Adamuz-Hinojosa, Oscar and Zanzi, Lanfranco and Sciancalepore, Vincenzo and Garcia-Saavedra, Andres and Costa-Perez, Xavier},
  title     = {{ORANUS}: Latency-tailored Orchestration via Stochastic Network Calculus in {6G O-RAN}},
  booktitle = {Proc. IEEE International Conference on Computer Communications (INFOCOM)},
  year      = {2024},
  pages     = {61--70}
}

@article{adamuz2025marea,
  author  = {Adamuz-Hinojosa, Oscar and Zanzi, Lanfranco and Sciancalepore, Vincenzo and Costa-Perez, Xavier},
  title   = {{MAREA}: A Delay-Aware Multi-time-Scale Radio Resource Orchestrator for {6G O-RAN}},
  journal = {IEEE Transactions on Communications},
  year    = {2025},
  volume  = {73},
  number  = {9},
  pages   = {7695--7710}
}

@article{giordani2020towards6g,
  author  = {Giordani, Marco and Polese, Michele and Mezzavilla, Marco and Rangan, Sundeep and Zorzi, Michele},
  title   = {Toward {6G} Networks: Use Cases and Technologies},
  journal = {IEEE Communications Magazine},
  year    = {2020},
  volume  = {58},
  number  = {3},
  pages   = {55--61}
}

@article{liu2022isacjsac,
  author  = {Liu, Fan and Masouros, Christos and Petropulu, Athina P. and Griffiths, Hugh and Han, T. X.},
  title   = {Joint Radar and Communication Design: Applications, State-of-the-Art, and the Road Ahead},
  journal = {IEEE Journal on Selected Areas in Communications},
  year    = {2022},
  volume  = {40},
  number  = {6},
  pages   = {1728--1767}
}

@article{kaushik2024isac6g,
  author  = {Kaushik, Aryan and Singh, Rohit and Shin, Wonjae},
  title   = {Integrated Sensing and Communication for {6G}: Recent Advances and Research Challenges},
  journal = {IEEE Communications Standards Magazine},
  year    = {2024},
  volume  = {8},
  number  = {2},
  pages   = {52--59}
}

@article{wu2024isac_interference,
  author  = {Wu, Huici and Wei, Zhiqing and Feng, Zhiyong},
  title   = {Interference Management for Integrated Sensing and Communication Systems},
  journal = {IEEE Internet of Things Journal},
  year    = {2024},
  volume  = {11},
  number  = {19},
  pages   = {31987--32002}
}

@article{polese2024openran,
  author  = {Polese, Michele and Dohler, Mischa and Melodia, Tommaso},
  title   = {Empowering the {6G} Cellular Architecture with Open {RAN}},
  journal = {IEEE Journal on Selected Areas in Communications},
  year    = {2024},
  volume  = {42},
  number  = {2},
  pages   = {245--259}
}

@article{meng2020pomdp_wireless,
  author  = {Meng, Xiang and Zeng, Yong},
  title   = {Optimal Resource Allocation in Wireless Systems with Delayed State Information},
  journal = {IEEE Transactions on Wireless Communications},
  year    = {2020},
  volume  = {19},
  number  = {5},
  pages   = {3497--3512}
}

@article{yang2023dt_control,
  author  = {Yang, Haoran and Xu, Wei and Huang, Yongming},
  title   = {Digital Twin-Enabled Control for Wireless Networks: Architecture and Applications},
  journal = {IEEE Wireless Communications},
  year    = {2023},
  volume  = {30},
  number  = {5},
  pages   = {58--65}
}

@article{park2021latency_ran,
  author  = {Park, Jinwook and Samarakoon, Sumudu and Bennis, Mehdi and Debbah, Merouane},
  title   = {Wireless Network Intelligence at the Edge: Latency, Reliability, and Scalability},
  journal = {IEEE Communications Magazine},
  year    = {2021},
  volume  = {59},
  number  = {7},
  pages   = {24--30}
}

\end{document}